\documentclass[
a4paper,fleqn,8pt]{article}
\usepackage{graphicx}
\usepackage{subfigure}
\usepackage{latexsym}
\usepackage{amsmath,amssymb}
\usepackage{dcolumn}

\begin{document}

\title{A Bound System in the Expanding Universe with Modified Holographic Ricci Dark Energy and Dark Matter}
\author {En-Kun Li,$^a$
Yu Zhang,$^{a,}$\thanks{Email: zhangyu\_128@126.com}
Jin-Ling Geng $^a$,
Peng-Fei Duan $^b$\\
$^a$ \it Faculty of Science, Kunming University of Science and Technology,\\ \it Kunming 650500, China.\\
$^b$ \it City College, Kunming University of Science and Technology,\\ \it Kunming 650051, China}

\date{}
\maketitle

\begin{abstract}
The evolution of a bound system in the expanding background has been investigated in this paper. The background is described by a FRW universe with the modified holographic dark energy model, whose equation of state parameter changes with time and can cross the phantom boundary. To study the evolution of the bound system, an interpolating metric is considered, and on this basis the geodesics of a test particle are given. The equation of motion and the effective potential are also derived from the geodesics. By studying the the effective potential and the evolution of the radius of a test particle in the bound system of the Milky Way galaxy, we have found that the galaxy would go through three stages: expands from a singular point; stays in a discoid for a period of time; big rip in the future. With the help of analysing the critical angular momentum, we find that the test particle needs less angular momentum to escape from the center mass as time passes.
\end{abstract}

\textbf{keywords:} bound system; expanding background; modified holographic dark energy

\section{Introduction}
\label{sec:int}

There is a great deal of observational evidence to prove that our universe is undergoing an accelerating expansion phase \cite{Riess,Perlmutter,Knop,Tonry,RiessB}. To explain this phenomenon, a large number of theoretical models have been put forward, and among these models the dark energy models have attracted a lot of attention \cite{Sahni,Padmanabhan,Peebles,SahniB,Copeland,Frieman,Li}. Although numerous dark energy models have been proposed, the nature of it is still very largely in the dark. Among these dark energy models, the so called holographic dark energy, which arises from the holographic principle \cite{Hooft,Susskind,Fischler,Bousso}, seems to provide a more simple and reasonable frame to investigate the problem of dark energy \cite{Setare,Sheykhi}. According to the holographic principle, the energy density of the holographic dark energy is written as $\rho_x= 3c^2 M_p^2 L^{-2}$, where $c$ is a numerical constant, and $M_p= 1/\sqrt{8\pi G}$ is the reduced Planck mass and $L$ is the largest IR cutoff. Until now, various types of IR cutoff have been considered to solve the dark energy problems, such as the Hubble radius \cite{Cohen,Horava,Thomas}, the particle horizon \cite{Fischler,BoussoB}, the future event horizon \cite{Huang,LiB}, the cosmological conformal time \cite{Cai,Wei}, or other generalized IR cutoff \cite{Chen,CaiB,Duran,Wang,Xu,XuB,XuC,Zhang}. Nowadays, the researches on the holographic dark energy have attracted so many scientists, and lots of remarkable works have been done in this field \cite{Bouhmadi,CaiC,del,Ghaffari,LiC,SetareC,SetareB,SetareD,SetareE,XuD}.

However, the current accelerating expansion can not be derived from the holographic dark energy models with Hubble horizon or particle horizon as the IR cutoff \cite{Li,Hsu}. Meanwhile, Setare et al. \cite{SetareB,SetareA} proved that the holographic dark energy model in the non-flat universe enclosed by the event horizon cannot cross the phantom divide line. To solve these problems, Gao, et al. \cite{Gao} raised a new holographic Ricci dark energy model, where the length scale is given by $L\sim|R|^{-1/2}$, and $R$ is the Ricci curvature scalar. Granda, et al. \cite{Granda,GrandaB} proposed a new form of IR cutoff for the holographic dark energy, which is a linear combination of $\dot{H}$ and $H^2$, i.e., $\rho_x\propto\alpha H^2+\beta \dot{H}$. This model could avoid the causality problem, which appears when using the event horizon area as the IR cutoff, meanwhile, the fine tuning problem and the coincidence problem can also be solved effectively. Recently, Chimento, et al. \cite{ChimentoA} proposed a modified form of holographic Ricci dark energy model, it is
\begin{eqnarray}
&\rho_x=\frac{2}{\alpha-\beta}\left(\dot{H}+\frac{3}{2}\alpha H^2\right),\label{eq:rhox}
\end{eqnarray}
where $\alpha$ and $\beta$ are free constants. After their work, the modified holographic Ricci dark energy (MHRDE) is studied in connection with the interacting dark matter in many papers, and more information can be found in \cite{Chattopadhyay,ChimentoB,ChimentoC,ChimentoD}.

In modern physics, the relation between cosmic expansion and local physics is a problem of principle in general relativity which still awaits a definitive answer \cite{Bars,Faraoni,GaoB,GaoC}. The physics of a bound system in the expanding universe is an immediate approach for this problem. A great deal of attention have been put on the effects of the universe's expansion on gravitationally bound system such as planetary, galaxies and cluster systems \cite{Bona,Bonnor,Cooperstock,Einstein,Gonzalez,Stefancic,Nesseris}. Among them, Nesseris, et al. \cite{Nesseris} have studied different bound systems in the phantom and quintessence universe, and they also gave the numerical reconstruction of the dissociating bound orbits. In their work, they found that the bound system got unbounded at the time when the minimum of the time-dependent effective potential disappeared. This is not the time when the phantom energy gravitational potential, which is due to the average $(\rho+3p)$, balance the attractive gravitational potential of the mass $M$ of the system. In the present paper, we would like to examine a bound system in an expanding universe filled with dark matter and MHRDE. In such a model, the equation of state parameter of MHRDE is time-depending.

This paper is organized as follows: In Sec. \ref{sec:MHRDE}, we give a brief review of the MHRDE model and the evolution of its equation of state parameter. The evolution of energy densities for dark matter and dark energy are studied next. In Sec. \ref{sec:bound}, using an interpolating metric, which is Schwarzschild like in small scales and a general time-depending FRW universe in large scales, we give the test particle's radial equation of motion. Based on this motion equation, the effective potential is given, too. Then, the bound system corresponding to the Milky Way galaxy is studied by discussing the evolution of the effective potential and the change of the circular orbits. The conclusions are given in Sec. \ref{sec:con}. In this work, we assume today's scale factor $a_0\rightarrow1$, far future $a\rightarrow \infty$, and the unit $8\pi G=1$.

\section{A brief review of the universe with the MHRDE model}
\label{sec:MHRDE}

Now, considering that there are only two components, i.e., the pressureless dark matter and the negative pressure dark energy. Then the Friedmann equation describing the evolution of the FRW universe can be written as
\begin{eqnarray}
  &H^2=\frac{1}{3}\left(\rho_m+\rho_x\right),\label{efr}
\end{eqnarray}
where $\rho_m$ and $\rho_x$ are the energy densities of dark matter and dark energy, $H=\dot{a}/a$ is the Hubble parameter and $a$ is the scale factor. The total energy density must satisfies the following conservation law:
\begin{eqnarray}
  &\dot{\rho}_m+\dot{\rho}_x+3H(\rho_m+\rho_x+p_x)=0.\label{etot}
\end{eqnarray}
Since we suppose that there is no interaction between dark matter and dark energy, the conservation law would be written in the following two separate equations
\begin{eqnarray}
  &\dot{\rho}_m+3H\rho_m=0,\label{ermc}\\
  &\dot{\rho}_x+3H(\rho_x+p_x)=0.\label{erxc}
\end{eqnarray}
Let's define that,
\begin{eqnarray}
  &h=\frac{H}{H_0},\quad \tilde{\rho}_m=\frac{\rho_m}{3H_0^2},\quad \tilde{\rho}_x=\frac{\rho_x}{3H_0^2},\label{edf}
\end{eqnarray}
where $H_0$ is the present value of the Hubble parameter. Substituting Eq. (\ref{edf}) into Eq. (\ref{ermc}), we obtain the equation for conformal density of dark matter as
\begin{eqnarray}
  &\tilde{\rho}_m=\Omega_{m0}a^{-3},\label{em}
\end{eqnarray}
where $\Omega_{m0}=\rho_{m0}/3H_0^2$ is the present density of dark matter. Using Eqs. (\ref{eq:rhox}), (\ref{edf}) and (\ref{em}), the Friedmann equation can be written in the following conformal form
\begin{eqnarray}
  &h^2=\Omega_{m0}a^{-3}+\frac{1}{3(\alpha-\beta)}\left(\frac{dh^2}{d\ln a}+3\alpha h^2\right).\label{eh1}
\end{eqnarray}
The general solution of the above differential equation is given by
\begin{eqnarray}
  &h^2=\frac{\alpha-\beta}{1-\beta}\Omega_{m0}a^{-3}+Ca^{-3\beta},\label{h2}
\end{eqnarray}
where $C$ is a constant and can be determined by the initial condition $h^2|_{x=0}=1$. Using the initial condition one can obtain $C =1 -\frac{ \alpha-\beta} {1-\beta} \Omega_{m0}$. Now, the conformal density of the MHRDE can be identified as
\begin{eqnarray}
  &\tilde{\rho}_x=\frac{\alpha-1}{1-\beta}\Omega_{m0}a^{-3}+\left(1-\frac{\alpha-\beta}{1-\beta}\Omega_{m0}\right)a^{-3\beta},\label{ex}
\end{eqnarray}
and the conformal Hubble parameter is
\begin{eqnarray}
  &h^2=&\Omega_{m0}a^{-3} +\frac{\alpha-1}{1-\beta}\Omega_{m0}a^{-3}\nonumber\\&&+\left(1-\frac{\alpha-\beta}{1-\beta}\Omega_{m0}\right)a^{-3\beta}.\label{eh}
\end{eqnarray}
Using Eqs. (\ref{erxc}) and (\ref{ex}), the equation of state parameter for the MHRDE can be written as
\begin{eqnarray}
  &w_x=-1+\frac{(\alpha-1)\Omega_{m0} +\beta\left[1-\beta-(\alpha-\beta)\Omega_{m0}\right]a^{3(1-\beta)}} {(\alpha-1)\Omega_{m0} +\left[1-\beta-(\alpha-\beta)\Omega_{m0}\right]a^{3(1-\beta)}}.
  \label{ewx}
\end{eqnarray}
From Eq. (\ref{ewx}), one can find that the equation of state parameter provides the possibility of transition from $w_x>-1$ to $w_x<-1$, which corresponds to the quintom model. If we consider a universe dominated by MHRDE, where the contribution from the DM is negligible, then Eq. (\ref{ewx}) becomes $w_x = -1+ \beta$. So if $\beta<0$, $w_x$ can cross the phantom divide.

However, from the above equations one can find that there are two free parameters that need to be determined. To determine one of them, we will take a special condition into consideration. The derivative of $H$ with respect to time is given by $\dot{H}=-\frac{3}{2}\frac{1+w_x+u}{1+u}H^2$, where $u=\tilde{\rho}_m/\tilde{\rho}_x$ is the ratio between dark matter and MHRDE, then Eq. (\ref{eh}) would be written in the following term
\begin{eqnarray}
  &h^2=\frac{1+u}{\alpha-\beta}\left(\alpha h^2-\frac{1+w_x+u}{1+u}h^2\right).
\end{eqnarray}
From this equation we obtain that the value of $\beta$ given in terms of the free parameter $\alpha$ as $\beta=u(1-\alpha)+(1+w_x)$. Taking the boundary conditions and the ratio between dark matter and MHRDE, i.e., $$ w_{x0} =-1,\quad u_0= \frac{\rho_{m0}}{\rho_{x0}} =\frac{\Omega_{m0}} {1-\Omega_{m0}}$$ into consideration, one can obtain
\begin{eqnarray}
  &\beta =\frac{\Omega_{m0}} {1-\Omega_{m0}} (1-\alpha).
  \label{ebeta}
\end{eqnarray}
From Eq. (\ref{ebeta}), it is easy to find that $\beta$ decreases as $\alpha$ increases. Now, the free parameters have been reduced to one, and for $\alpha=1$ we have $\beta=0$.
\begin{figure}[!htb]
\centering
  \includegraphics[scale=0.5]{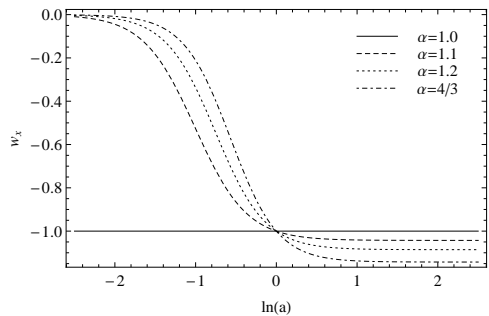}
\caption{The evolution of the state parameter of MHRDE with respect to $\ln a$, we choose $\Omega_{m0}=0.3$.}
\label{p0}
\end{figure}
\begin{figure}[!htb]
\centering
  \includegraphics[scale=0.5]{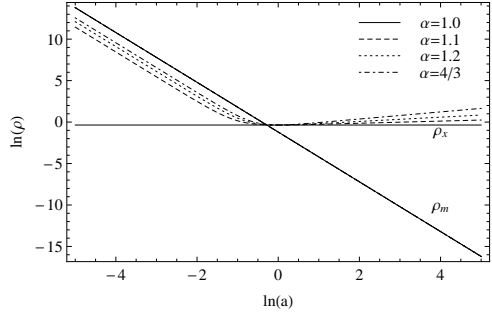}
\caption{The evolution of the energy density of dark matter and MHRDE, we choose $\Omega_{m0}=0.3$.}
\label{p1}
\end{figure}

The evolutions of the state parameter of MHRDE and the densities of dark matter and MHRDE are plotted in Figs. \ref{p0} and \ref{p1}, respectively. Combining the two figures, one can easily find that the future of the universe would be MHRDE dominated. When $\alpha=1$, as shown in Fig. \ref{p0}, the equation of state parameter of MHRDE is $w_x=-1$, and as shown in Fig. \ref{p1}, its energy density is a constant, which indicates that the MHRDE is $\Lambda$CDM like when $\alpha=1$.


\section{Evolution of the bound system}
\label{sec:bound}

In order to study the bound system in the expanding universe, one needs the geodesics of a test particle in the expanding universe. Now, let's consider an appropriate metric that satisfies the static Schwarzschild metric at small distance and a time-depending FRW space time at large distances. Following most authors \cite{Einstein,Nesseris,McVittie}, we consider such an interpolating metric, which is under the Newtonian limit (weak field, low velocities), takeing the form \cite{Nesseris}:
\begin{eqnarray}
  &ds^2=&\left(1-\frac{2GM}{a(t)\rho}\right)\cdot dt^2-a(t)^2\cdot[d\rho^2\nonumber\\ &&+\rho^2\cdot(d\theta^2+\sin^2\theta d\varphi^2)],\label{eq:metric}
\end{eqnarray}
where $\rho$ is the comoving radial coordinate. Using $r= a(t) \rho$, one can obtain the geodesics
\begin{eqnarray}
  &-\left(\ddot{r}-\frac{\ddot{a}}{a}r\right)-\frac{GM}{r^2}+r\dot{\varphi}^2=0,\label{egr}\\
  &r^2\dot{\varphi}=L,\label{egl}
\end{eqnarray}
where $L$ is the constant angular momentum per unit mass. Then, the radial equation of motion for a test particle in the Newtonian limit would be
\begin{eqnarray}
  &\ddot{r}=-\frac{GM}{r^2}+\frac{L^2}{r^3}+\frac{\ddot{a}}{a}r.
  \label{ed2tr}
\end{eqnarray}
From this equation one can obtain the effective potential, which determines the dynamics of the bound system, by $\ddot{r}=-\partial V/\partial r$. Integrating it with respect to $r$ and ignoring the arbitrary integration function of time, one can obtain the effective potential
\begin{eqnarray}
  &V_{eff}(t,r)=-\frac{GM}{r}+\frac{L^2}{2r^2}-\frac{\ddot{a}}{2a}r^2.
  \label{eV}
\end{eqnarray}

Here, taking $\frac{d}{dt} =H \frac{d} {d\ln a}$ into consideration, we can obtain
\begin{eqnarray}
  &\dot{r}=\frac{dr}{dt}=Hr',
\end{eqnarray}
where $r'=dr/d\ln a$, then for $\ddot{r}$ one has
\begin{eqnarray}
  &\ddot{r}=\frac{d\dot{r}}{dt}=\frac{d(Hr')}{d\ln a}\frac{d\ln a}{dt}=H^2r''+\frac{1}{2}(H^{2})'r'.
  \label{ed2xr}
\end{eqnarray}
We also take $\ddot{a}/a =\dot{H} +H^2$ into account and use Eq. (\ref{ed2xr}), then, Eq. (\ref{ed2tr}) would be
\begin{eqnarray}
  &H^2r''+\frac{1}{2}(H^2)'r'=&-\frac{GM}{r^2}+\frac{L^2}{r^3}\nonumber\\&&+[\frac{1}{2}(H^2)'+H^2]\cdot r.
  \label{eHr}
\end{eqnarray}
Similarly, one can obtain the effective potential in terms of $H$ as
\begin{eqnarray}
  &V_{eff}=-\frac{GM}{r}+\frac{L^2}{2r^2}-\frac{1}{2}[\frac{1}{2}(H^2)'+H^2]r^2.
  \label{eHV}
\end{eqnarray}

In what follows, we will focus on the evolution of a two body bound system in the expanding unverse. Let us assume that at the present time, i.e., $a=a_0$, a test particle moves on a circular orbit with radius $r_0$, and $\dot{\varphi}(t_0)^2 =\omega_0^2 =GM/r_0^3$. Taking the following conformal forms into consideration
\begin{eqnarray}
  &\omega_0^2=\frac{GM}{r_0^3H_0^2}, \quad r=r/r_0, \quad V_{eff}=V_{eff}/(H_0^2r_0^2),
\end{eqnarray}
Eqs. (\ref{eHr}) and (\ref{eHV}) could be written in the following forms
\begin{eqnarray}
  &h^2r''+\frac{1}{2}(h^2)'r'=-\frac{\omega_0^2}{r^2}+\frac{\omega_0^2}{r^3}+\lambda(a)^2 r,\label{ehr}\\
  &V_{eff}=-\frac{\omega_0^2}{r}+\frac{\omega_0^2}{2r^2}-\frac{1}{2}\lambda(a)^2 r^2,\label{ehV}
\end{eqnarray}
where
\begin{eqnarray}
  &\lambda(a)^2=\frac{1}{2}(h^2)'+h^2.\label{ela}
\end{eqnarray}
By analysing the effective potential, one could find the radius of the circular orbit is given by the minimum $r_{min}(a)$ of the effective potential. From Eq. (\ref{ehV}), we know that the location of $r_{min}(a)$ is depending on the scale factor. Radius of the circular orbit is given by solving the following equation
\begin{eqnarray}
  &\frac{\lambda(a)^2}{\omega_0^2}r_{min}^4-r_{min}+1=0.\label{ermin}
\end{eqnarray}
In fact, Eq. (\ref{ermin}) has a solution only for $\lambda(a)^2 \leq 27\omega_0^2/256$ \cite{Nesseris}. Therefore, when $\lambda(a)^2= 27\omega_0^2/256$, the minimum of the effective potential disappears, then, the system becomes unbound. Here, we define that the solution of equation $\lambda(a)^2= 27\omega_0^2/256$ as $a=a_{rip}$, which means the big rip occurs at the trip time $a_{rip}$.

\begin{figure}[!htb]
\centering
  \includegraphics[scale=0.5]{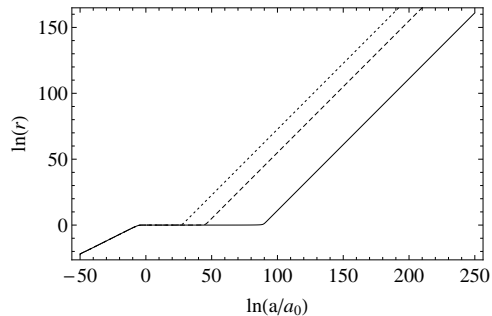}
\caption{The numerical evolution of the radius for the Milky Way galaxy is plotted under different cases: continuous line for $\alpha=1.1$, dashed line for $\alpha=1.2$ and dotted line for $\alpha=4/3$. Here we choose $\Omega_{m0}=0.3$.}
\label{p2}
\end{figure}

\begin{figure}[!htb]
\centering
  \includegraphics[scale=0.4]{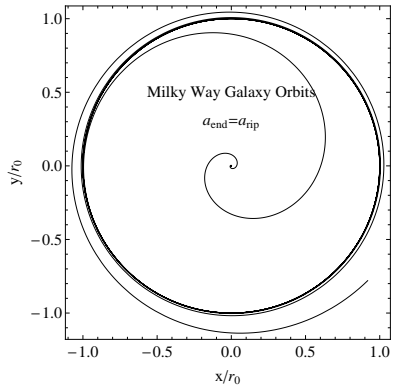}
\caption{The evolution of the system of the Milky Way galaxy until the potential minimum disappears. Here, we choose $\alpha=1.2$ and $\Omega_{m0}=0.3$.}
\label{p3}
\end{figure}

To investigate the future radial evolution of a bound gravitational system, we will take a specific bound system --- the Milky Way galasy ($M= 2 \times 10^{45} gr$, $r_0 = 5 \times 10^{22} cm$, $\omega_0 = 182$) --- into consideration. The numerical evolution of the radius for different $\alpha$ is shown in Fig. \ref{p2}. As shown in Fig. \ref{p2}, it is easy to find that all the curves have two turning point, one occurs at small $a$, which means the early universe, and another occurs at big $a$, which means the future universe. According to Fig. \ref{p2}, we find that $r<r_0$ occurs before the first turning point, and $r>r_0$ occurs after the second turning point. From this we can say that the Milky Way galaxy would go through three stages: the first stage is the expanding stage: from a very small radius to today's size; the second is the steady stage: the radius changes little and the galaxy stays in a discoid; the third is the rip stage: the radius of the galaxy becomes bigger and bigger that the galaxy would be ripped. From Fig. \ref{p2}, one can also find that as $\alpha$ increases the values of $a_{rip}$ decreases, which means that the big rip occurs more early. The two body bound system corresponding to the Milky Way galaxy is plotted in Fig. \ref{p3}. From Fig. \ref{p3}, it is more easy to find that the galaxy comes from a singular point, stays in discoid for a period of time, and will be ripped finally.

\begin{figure}[!htb]
\centering
  \includegraphics[scale=0.5]{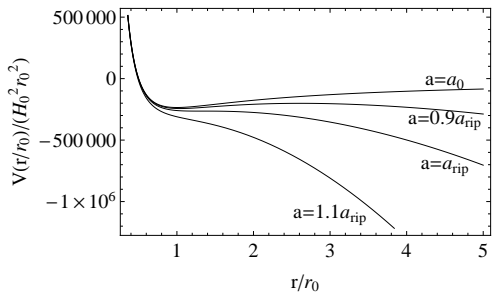}
\caption{The evolution of the effective potential corresponding to different scale factor $a$ with respect to the dimensionless radius $r/r_0$, and we choose $\alpha=1.2$, $\Omega_{m0}=0.3$.}
\label{p4}
\end{figure}

The evolution of the effective potential around the trip time $a_{rip}$ is shown in Fig. \ref{p4}. As a contrast, the effective potential of the present time is also plotted in Fig. \ref{p4}. As we know, the stability of a bound system depends on whether the effective potential has a minimum. From Fig. \ref{p4}, one can find that when $a<a_{rip}$, as the radius $r$ increases the effective potential curves decreases to a minimum first, then increases to a maximum and finally decreases to $-\infty$, and its maximum decreases as the scale factor increases. Then we know that the effective potential curves have a minimum, which indicates the possibility of bound orbits for the test particle. However, as we can see from Fig. \ref{p4}, when $a=a_{rip}$, the effective potential curve has only one inflection point and when $a>a_{rip}$ there is no inflection point. Then, we can say that when the scale factor $a\geq a_{rip}$, the gravitational force can not bound the test particle any longer, they will fly to distant area under the negative pressure force of the MHRDE.

\begin{figure}[!htb]
\centering
  \includegraphics[scale=0.5]{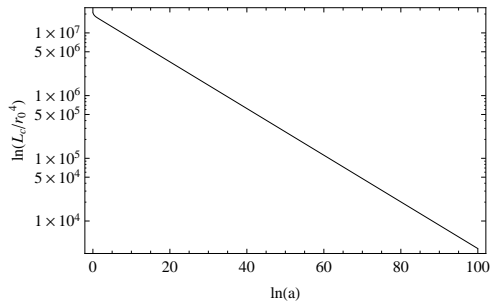}
\caption{The critical angular momentum $\tilde{L}_c$ with respect to $\ln a$. Here, we choose $\alpha=1.2$ and $\Omega_{m0}=0.3$.}
\label{p5}
\end{figure}

We have studied the test particle at a circular orbit above, where we assume that $\dot{\varphi}(t_0)^2 =\omega_0^2 =GM/r_0^3$. Now, let's consider a different case that the test particle has different angular momentums. Eq. (\ref{ehV}) could be written in the following form
\begin{eqnarray}
  &V_{eff}=-\frac{\omega_0^2}{r}+\frac{\tilde{L}^2}{2r^2}-\frac{1}{2}\lambda(a)^2 r^2,\label{ehVL}
\end{eqnarray}
where $\tilde{L}^2 =L^2/(H_0^2 r_0^4)$. From Eq. (\ref{ehVL}) one can find that when $\tilde{L}^2= \tilde{L}_c^2 =\frac{3}{4} \omega_0^2 [\omega_0^2/ (4\lambda(a)^2)]^{ \frac{1}{3}}$, the effective potential curves have no extremum. The critical angular momentum $\tilde{L}_c$ with respect to $\ln a$ is plotted in Fig. \ref{p5}. From Fig. \ref{p5}, one can obtain that the critical angular momentum decreases as $\ln a$ increases, which means with the passage of time, the test particle would need less angular momentum to escape the gravitational force from the center mass.

\begin{figure}[!htb]
\centering
  \includegraphics[scale=0.7]{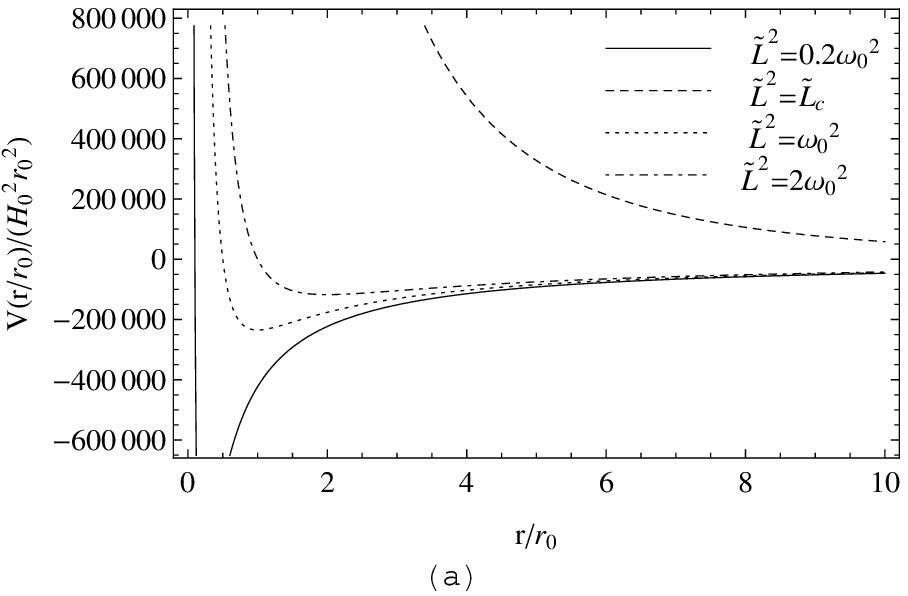}
  \includegraphics[scale=0.7]{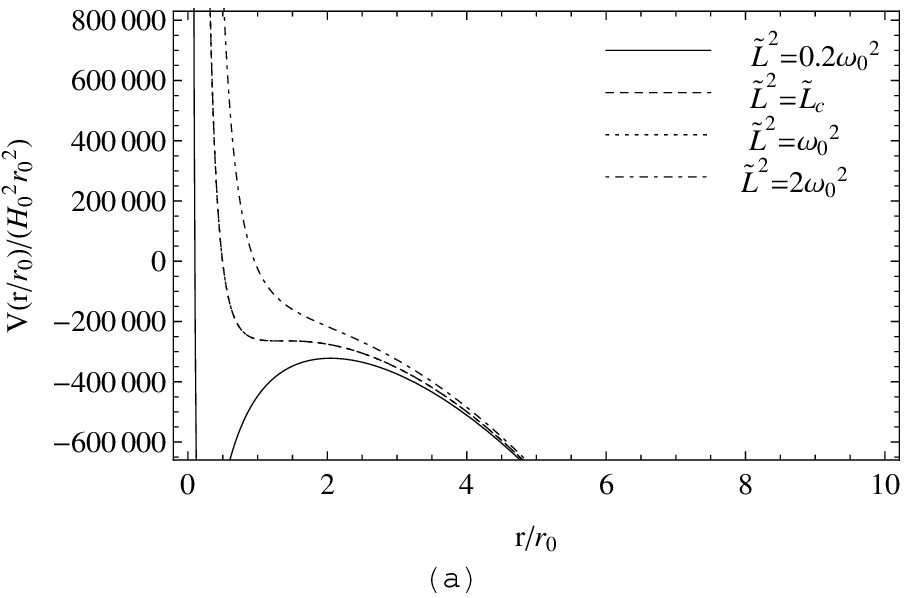}
\caption{The evolution of the effective potential with respect to $\ln a$ under different angular momentum. In figure (a), we choose the scale factor as $a=a_0$ and in figure (b) we choose $a=a_{rip}$. Here, we choose $\alpha=1.2$ and $\Omega_{m0}=0.3$.}
\label{p6}
\end{figure}

In order to make this more clear, the effective potential curves for the test particle with different angular momentums are shown in Fig. \ref{p6}. From Fig. \ref{p6} (a), one can find that the critical angular momentum $\tilde{L}$ is more big than $\omega_0$, i.e., $\tilde{L}_c^2>>\omega_0^2$. Fig. \ref{p6} (b) shows that $\tilde{L}_c^2=\omega_0^2$. As we all know, when the effective potential has no extremum, the test particle could not be bounded together. Then, with the help of the two graphics in Fig. \ref{p6}, we can say that the test particle would need less angular momentum to escape the immense pull from the center mass with the passing of time.



\section{Conclusion}
\label{sec:con}

In this paper, we have investigated the evolution of a bound system in the expanding universe, the universe is described by the FRW universe with only two cosmic components: dark matter and dark energy. In the present paper the dark energy is considered as the MHRDE, whose density is $\rho_x=\frac{2}{\alpha-\beta} (\dot{H}+\frac{3}{2}H^2)$ and the equation of state parameter of it is time-depending. From Sec. \ref{sec:MHRDE} one can find that the MHRDE behaves like the quintom.

To investigate the bound system in the expanding universe, we have considered an interpolating metric which can reduce to the static Schwarzschild metric at small scales but FRW metric at large scales. The geodesics of a test particle are derived by this metric and using the geodesics we get the radial equation of motion for a two body bound system. The test particle's motion in the Milky Way galaxy is examined by analysing the equation of motion and the effective potential. We have found that the the galaxy would go through three stages: the first stage is expanding from a singular point, then it will stay in a discoid for a period of time, and it will be ripped owing to the negative pressure of the MHRDE finally.

Moreover, by studying the evolution of the critical angular momentum with respect to the scale factor $a$, we find that with the passage of time, the test particle would need less angular momentum to escape the immense pull of the center mass.

\section*{Acknowledgments}

This work was supported in part by the National Natural Science Foundation of China (Grants No. 11347101, No. 11405076), the Science Foundation of the Education Department of Yunnan Province (Grant No. 2014Y066), and the Talent Cultivation Foundation of Kunming University of Science and Technology (Grants No. KKSY201207053, No. KKSY201356060). Yu Zhang would like to acknowledge the support of the working Funds of the Introduced High-level Talents of Yunnan Province from the Department of Human Resources and Social Security of Yunnan Province.

\end{document}